\documentclass[aps,prb,twocolumn,superscriptaddress,showpacs]{revtex4}

\usepackage{graphicx}
\begin{document}

\hyphenation{TiOCl TiOBr}


\title{Unusual quasi-one-dimensional electron dispersions in the spin-1/2 quantum magnet TiOCl}



\author{M.~Hoinkis}
\affiliation{Experimentalphysik II, Universit\"at Augsburg,
D-86135 Augsburg, Germany}

\author{M.~Sing}
\affiliation{Experimentalphysik II, Universit\"at Augsburg,
D-86135 Augsburg, Germany} \affiliation{Experimentelle Physik 4,
Universit\"at W\"urzburg, D-97074 W\"urzburg, Germany}

\author{J.~Sch\"afer}
\author{M.~Klemm}
\author{S.~Horn}
\affiliation{Experimentalphysik II, Universit\"at Augsburg,
D-86135 Augsburg, Germany}


\author{H.~Benthien}
\affiliation{Fachbereich Physik, Philipps-Universit\"at, D-35032
Marburg, Germany}

\author{E.~Jeckelmann}
\affiliation{Institut f\"ur Physik, Johannes
Gutenberg-Universit\"at, D-55099 Mainz, Germany}

\author{T.~Saha-Dasgupta}
\affiliation{S. N. Bose Centre for Basic Sciences, Kolkata 700098,
India}

\author{L.~Pisani}
\author{R.~Valent\'\i}
\affiliation{Institut f{\"u}r Theoretische Physik, Universit{\"a}t
Frankfurt, D-60054 Frankfurt, Germany}

\author{R.~Claessen}
\email[]{claessen@physik.uni-wuerzburg.de}
\affiliation{Experimentalphysik II, Universit\"at Augsburg,
D-86135 Augsburg, Germany} \affiliation{Experimentelle Physik 4,
Universit\"at W\"urzburg, D-97074 W\"urzburg, Germany}

\date{\today}

\begin{abstract}
We have studied the electronic structure of the spin-1/2 quantum
magnet TiOCl by polarization-dependent momentum-resolved
photoelectron spectroscopy. From that, we confirm the
quasi-one-dimensional nature of the electronic structure along the
crystallographic $b$-axis and find no evidence for sizable
phonon-induced orbital fluctuations as origin for the
non-canonical phenomenology of the spin-Peierls transition in this
compound. A comparison of the experimental data to our own LDA+U
and Hubbard model calculations reveals a striking lack of
understanding regarding the quasi-one-dimensional electron
dispersions in the normal state of this compound.
\end{abstract}


\maketitle

\section{Introduction}
%
%
Strongly correlated low-dimensional quantum \mbox{spin-1/2}
systems have attracted much attention during the last years since
they can serve as a testing ground for theo\-re\-ti\-cal concepts
which are believed to play a key role in understanding materials
of paramount technological potential like the high-$T_c$
superconductors or the CMR (colossal magnetoresistance)
manganites.\cite{Imada99,Dagotto99} This applies in particular to
one-dimensional systems for which exactly solvable many-body
models exist.\cite{Voit95} Meanwhile the narrowed view on the
isolated subsystems, e.g. the electron system, has given way to a
synopsis of the complex interplay between {\it all} relevant
degrees of freedom, i.e. charge, spin, orbitals, and lattice. In
this perspective, the layered Mott insulator TiOCl has recently
aroused a lot of interest.

TiOCl crystallizes in an orthorhombic quasi-two-dimensional
structure of the FeOCl type, where buckled bilayers of Ti-O are
separated by Cl ions.\cite{Schaefer58} The bilayers, which are
stacked along the crystallographic $c$-axis, only weakly interact
through van der Waals forces. In a local picture, the electronic
structure is determined by the octahedral coordination of the Ti
ion in a 3$d^1$ configuration. The strongly distorted octahedra
are formed by four O and two Cl ions, and share corners along the
$a$-axis and edges along the $b$-axis. Thus, the low-lying charge
excitations occur within the Ti~3$d$~$t_{2g}$ triplet. This simple
picture is essentially confirmed by LDA+U
calculations,\cite{Seidel03,Saha-Dasgupta04} which identify the
$d_{xy}$ derived band as slightly split off from the bands with
$d_{xz}$ and $d_{yz}$ character.\cite{footnote-xyz} Both the LDA+U
results as well as the observation of a Bonner-Fisher-type
magnetic susceptibility at high temperatures indicate the
existence of spin-1/2 Heisenberg chains in this material, mediated
by direct exchange along the $b$-axis ($J \approx
660$\,K).\cite{Seidel03} Correspondingly, the sudden drop of the
susceptibility at $T_{c_1}=67$\,K to almost zero and a kink
anomaly at $T_{c_2}=91$\,K have been discussed in terms of an
unusual first-order spin-Peierls transition and possibly some kind
of precursor transition, respectively. While the dimerized nature
of the ground state indeed could be proven by x-ray
diffraction,\cite{Shaz05} the results from magnetic
resonance,\cite{Imai03,Kataev03} Raman\cite{Lemmens04} and
infrared spectroscopy,\cite{Caimi04} and specific heat
measurements\cite{Hemberger05} point to the importance of strong
spin and/or orbital fluctuations in the high-temperature phase up
to 130\,K. In partial contradiction, there is recent evidence from
cluster calculations in connection with polarization-dependent
optical data that the orbital degrees of freedom are actually
quenched.\cite{Ruckamp05} Based on Ginzburg-Landau arguments it
was concluded that the frustration of the interchain interactions
in the bilayers give rise to incommensurate order in the
intermediate phase, which commensurately locks in below $T_{c_1}$.
Such behavior has indeed been observed in the related compound
TiOBr.\cite{vanSmaalen05} Still, a complete understanding of the
relevant competing interactions in this compound demands for a
thorough investigation of the electronic properties in the normal
state. This seems particularly interesting under the perspective
of doping charge carriers into the system to drive the chains
metallic and possibly even superconducting.\cite{Beynon93,Craco04}

%
%

In this paper we focus on the normal state electronic properties
of TiOCl at room-temperature and slightly above, {\it i.e.} well
above the transition temperatures $T_{c_1}$ and $T_{c_2}$. To this
end we performed polarization dependent angle-resolved
photoemission measurements (ARPES) which provide momentum-resolved
information on the electronic structure. The experiments are
complemented by LDA+U and Hubbard model calculations. Our
photoemission data confirm the one-dimensionality of the
electronic structure and give no evidence for any significant
(phonon-induced) admixtures of $d_{xz}$ and $d_{yz}$ orbitals to
the ground state, implying that orbital fluctuations play no role
for the low-temperature physics of TiOCl. Nevertheless, the
detailed comparison of the experimental results to our electronic
structure calculations as well as to recent LDA+DMFT calculations
show that even the normal state properties of TiOCl are far from
being fully understood.

\section{Technical details}
\subsection{Experimental}
%
%

\begin{figure}
\includegraphics[width=8.2cm]{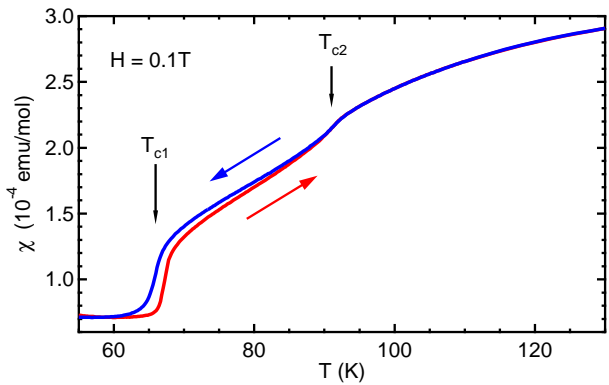}
\caption{\label{Figure1}(Color online) Magnetic susceptibility of
TiOCl measured by SQUID magnetometry.}
\end{figure}

Single crystals of TiOCl were prepared by chemical vapor transport
from TiCl$_3$ and TiO$_2$.\cite{Schaefer58} The samples were
characterized by x-ray diffraction, specific heat, magnetic
susceptibility, and electron spin resonance measurements. The high
quality of our crystals can be directly inferred from the
susceptibility data displayed in Fig.~\ref{Figure1}, in which a
pronounced hysteresis at $T_{c_1}$ signals the first-order
character of the lower transition.

Photoemission spectroscopy has been performed mainly at our
homelab using He I (21.2\,eV) and Al K$_{\alpha}$ (1486.6\,eV)
radiation and an Omicron EA 125 HR electron energy analyzer. For
polarization-dependent measurements we utilized an Omicron AR 65
spectrometer equipped with a He discharge lamp and a rotatable
polarizer. Additional polarization experiments with linearly
polarized synchrotron light were performed at the SIS beamline of
the Swiss Light Source at the Paul-Scherrer-Institute (Villigen,
Switzerland) using a Scienta SES 100 analyzer. The energy and
angular resolution amounted for the homelab experiments to 60\,meV
and $\pm1^\circ$, respectively, and to 80\,meV and $\pm 0.2^\circ$
in the SLS experiments. The layered structure of TiOCl facilitates
easy surface preparation by {\it in situ} crystal cleavage. The
resulting surfaces were clean and atomically long-range ordered as
evidenced by x-ray induced photoemission (XPS) and low-energy
electron diffraction (LEED), respectively. Because TiOCl is an
insulator, all photoemission data were taken at room temperature
or above in order to minimize sample charging. From systematic
temperature-variations we found that the charging is almost
negligible at and above $\approx 370$\,K. In this situation the
maximum of the Ti~$3d$-peak (see below) is located at 1.45\,eV
below the experimental chemical potential $\mu_{exp}$, which
corresponds to the Fermi edge position of a silver foil. Spectra
measured at lower temperatures have been aligned accordingly.

\subsection{LDA+U calculations}

On the theoretical side, we have determined the electronic
structure of TiOCl by performing density functional theory (DFT)
calculations in the generalized gradient approximation
(GGA)\cite{Perdew96} and in the so-called LDA\,(GGA)+U
approximation\cite{Anisimov93} using the full-potential linearized
augmented plane-wave code WIEN2k.\cite{wien2k} Since the LDA+U
calculations are performed on a spin-polarized state, we
considered a ferromagnetic alignment of the Ti spins (FM) as well
as an antiferromagnetic arrangement of the Ti spins along the $b$
direction (AFM). The latter was found to be lower in total energy
compared to the FM state.\cite{Saha-Dasgupta04} We also performed
LDA+U calculations for the low-temperature crystal structure of
TiOCl\cite{Shaz05} in an antiferromagnetically spin-polarized
state. In all calculations we used $RK_{\mathrm{max}}=6$ and
$56$\,k irreducible points for Brillouin-zone integrations. The
values for the onsite Coulomb repulsion $U$ as well as onsite
exchange $J_0$ were taken to be $3.3$\,eV and $1$\,eV,
respectively,\cite{Seidel03,Saha-Dasgupta04} which accounts well
for the intersite chain exchange constant derived from the
magnetic susceptibility.

\subsection{Dynamical density-matrix renormalisation group calculations}

To better account for correlation effects beyond the LDA+U
approach we have also determined the one-particle spectral
function of the one-dimensional (1D) Hubbard model at half
band-filling using the dynamical density-matrix renormalisation
group (DDMRG) method. \cite{Jeckelmann02, Benthien04} The
calculations were performed on 32-site chains with open boundaries
and we kept up to 200 density-matrix eigenstates. The broadening
of the spectra is $\eta/t$\,=\,0.2.

\section{Results and discussion}
\subsection{Valence density of states}

\begin{figure}
\includegraphics[width=8.2cm]{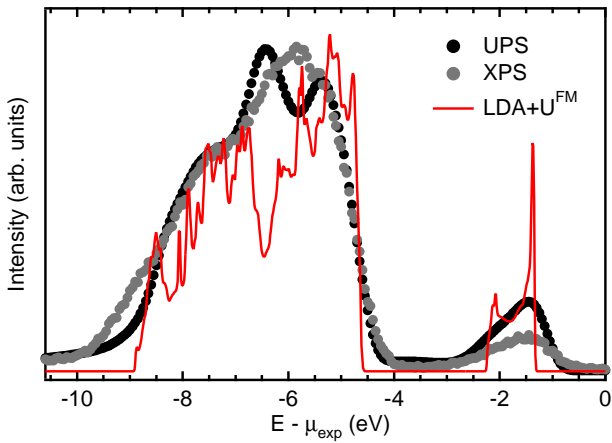}
\caption{\label{Figure2}(Color online) Angle-integrated
photoemission spectra and LDA+U density of states for a
ferromagnetic spin-polarized state . The UPS and XPS spectra were
measured at photon energies of 21.2\,eV and 1486.6\,eV,
respectively.}
\end{figure}

Figure \ref{Figure2} shows angle-integrated photoemission spectra
measured with 21.2\,eV (UPS) and 1486.6\,eV (XPS) photons. Both
spectra consist of two well separated regions. The small intensity
differences between the UPS and XPS data are due to different
photoemission cross sections. Also shown is the occupied part of
the density of states (DOS) obtained from the LDA+U calculation
for the room-temperature structure assuming ferromagnetic spin
alignment (LDA+U$^{\mathrm{FM}}$). From the comparison of theory
and experiment we identify the low-binding energy peak as Ti
$3d$-like, whereas the states between $-9$\,eV and $-4$\,eV are
predominantly derived from Cl $3p$ and O $2p$ levels. The overall
shape of the theoretical DOS and the experimental spectra are
rather similar. Note, however, that the LDA+U DOS is shifted to
higher binding energy in order to align the theoretical and
experimental Ti $3d$ peaks. With this alignment the relative
separation between Ti $3d$ and the ligand $p$ states still appears
$\approx 0.5$\,eV too small. The value of 2\,eV for the
correlation gap as determined from optical
spectroscopy\cite{Maule88, Grueninger04} is in reasonable
agreement with our photoemission data and the LDA+U calculations
(not seen here, as the figure displays only the occupied DOS).

\begin{figure}
\includegraphics[width=8.2cm]{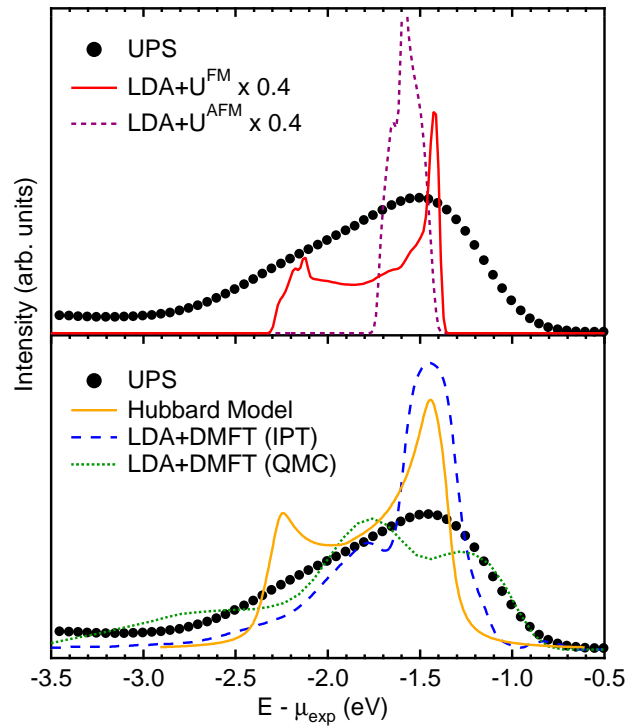}
\caption{\label{Figure3}(Color online) Ti $3d$ part of the
angle-integrated photoemission spectrum ($h\nu=21.2$\,eV) compared
to the density of states obtained from various calculations. Upper
panel: LDA+U with ferromagnetic and antiferromagnetic spin order
($U=3.3$\,eV). Lower panel: 1D Hubbard model, LDA+DMFT using
iterative perturbation theory (Ref.~\onlinecite{Craco04}), and
LDA+DMFT using quantum Monte Carlo
(Ref.~\onlinecite{Saha-Dasgupta05}). All spectra are normalized to
same integrated area.}
\end{figure}

In the following we focus on the Ti $3d$ part near the chemical
potential, which is shown as a blow-up in Fig.~\ref{Figure3}.
According to the LDA+U calculations\cite{Seidel03,Saha-Dasgupta04}
it has exclusively $3d_{xy}$ orbital character. As seen in the
upper panel of Fig.~\ref{Figure3}, the experimental $3d$ spectrum
has a similar asymmetry as the LDA+U$^{\mathrm{FM}}$ DOS but is
much broader. We emphasize that the experimental width is
perfectly reproducible and {\it not} due to instrumental
broadening, which is much smaller. A comparison to the LDA+U DOS
for {\it antiferromagnetic} spin configuration (which seems the
more natural choice considering the Bonner-Fisher-type
susceptibility) yields an even stronger disagreement.

We note that LDA+U accounts for the onsite Coulomb interaction
only in a mean-field way and is thus effectively still a
one-electron theory for static ordered systems. One may thus
speculate that the origin for the conflicting $3d$ widths lies in
pronounced electronic correlation effects and/or fluctuations of
spin-Peierls or orbital nature beyond the scope of LDA+U.
Electronic correlation effects can in principle be accounted for
by a combination of LDA and dynamical mean field theory (LDA+DMFT)
or by suitable many-body models. The lower panel of
Fig.~\ref{Figure3} shows a comparison of the experimental Ti $3d$
spectrum to two different LDA+DMFT calculations and the DOS of the
half-filled 1D single-band Hubbard model calculated using the
DDMRG method (using $U=3.3$\,eV and a transfer integral of
$t=0.23$\,eV, corresponding to the LDA+U calculations). The
LDA+DMFT calculations are taken from Refs.~\onlinecite{Craco04}
and \onlinecite{Saha-Dasgupta05} and used different impurity
solvers and basis sets. While the LDA+DMFT curves indeed display a
broadening much closer to that of the experiment, none of these
curves can sufficiently explain the shape of the experimental
data. The striking disagreement between the two LDA+DMFT
calculations should probably be traced back to the fact that,
contrary to the multi-orbital QMC solver,  the IPT solver is quite
uncertain for anisotropic multi-band
problems,\cite{Lichtenstein98} as it is the case for TiOCl. Also
the 1D Hubbard model does not reproduce the bandwidth and the
detailed spectral shape of our photoemission data.

\subsection{Dispersions and anisotropy of the electronic structure}
\begin{figure}
\includegraphics[width=8.2cm]{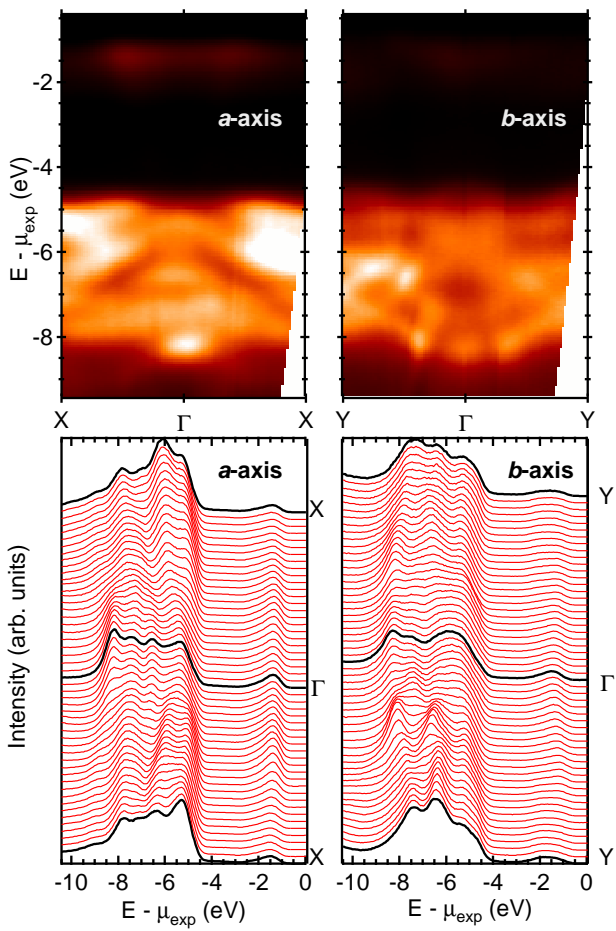}
\caption{\label{Figure4}(Color online) ARPES intensity plots
$I(\textbf{k},E)$ and EDCs along the crystallographic axes $a$ and
$b$, corresponding to the X$\Gamma$X and Y$\Gamma$Y lines in the
orthorhombic Brillouin zone.}
\end{figure}

We now turn to our angle-resolved photoemission data, which are
shown in Fig.~\ref{Figure4} in a broad energy range for the two
crystallographic directions $a$ and $b$. Both data sets are
displayed in two alternative ways: as intensity plots
$I(\textbf{k},E)$, and as energy distribution curves (EDCs). As
seen, there is well-pronounced dispersion particularly in the Cl
$3p$/O $2p$ part of the spectra between $-9$\,eV and $-4$\,eV.
This behavior and the clear symmetry of the dispersions with
respect to the $\Gamma$ point is indicative of the good crystal
surface quality.

\begin{figure}
\includegraphics[width=8.2cm]{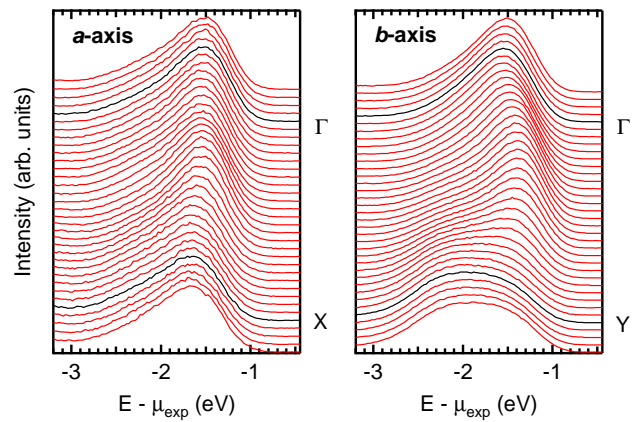}
\caption{\label{Figure5}(Color online) Ti $3d$ part of the EDCs
along the crystallographic axes $a$ and $b$.}
\end{figure}

\begin{figure}
\includegraphics[width=8.2cm]{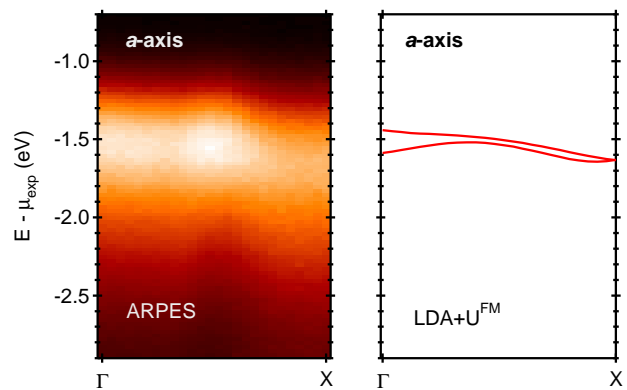}
\caption{\label{Figure6}(Color online) ARPES intensity plot
$I(\textbf{k},E)$ and LDA+U$^{\mathrm{FM}}$ bands along the
crystallographic $a$-axis.}
\end{figure}

Figure~\ref{Figure5} shows an enlargement of the Ti $3d$ ARPES
spectra. From a comparison of the bare EDCs one can easily see
that the spectral changes along the $b$-axis are considerably
stronger than along the $a$-axis. This observation confirms the
quasi-one-dimensional electronic (and magnetic) nature of TiOCl
and corroborates the LDA+U prediction\cite{Seidel03} that \mbox{1D
chains} are formed by direct hopping via Ti $3d_{xy}$ orbitals
along the $b$ direction. We note, however, that the dispersion
along the $a$-axis is still finite as seen from a closer
inspection of the $k$-dependence of the peak maximum, which moves
to slightly higher binding energy from $\Gamma$ to X. The size of
the $a$-axis dispersion is directly related to the interchain
coupling, which may play an important role for the complex
spin-Peierls transition behavior.\cite{Ruckamp05} A comparison of
the experimental $a$-axis dispersion to that of the
LDA+U$^{\mathrm{FM}}$ calculation is displayed in
Fig.~\ref{Figure6} and yields qualitative agreement concerning
size and direction of the dispersion. Note again, however, that
the LDA+U bands had to be shifted in energy to give a good match
to the ARPES data.

\begin{figure*}
\includegraphics[width=17.5cm]{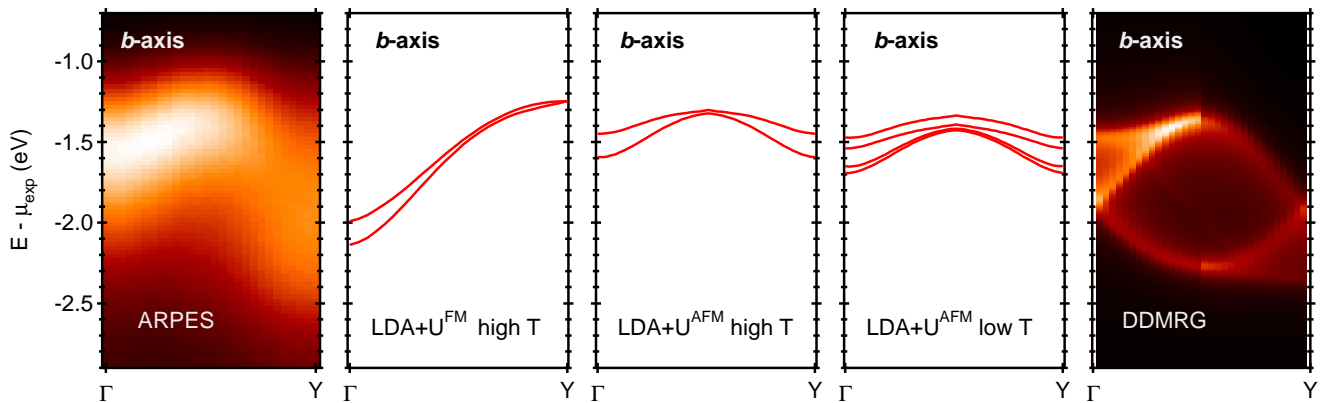}
\caption{\label{Figure7}(Color online) ARPES intensity plot
$I(\textbf{k},E)$, LDA+U$^{\mathrm{FM/AFM}}$ bands calculated for
the high-temperature structure ($U=3.3$\,eV, $J_0=1$\,eV),
LDA+U$^{\mathrm{AFM}}$ bands calculated for the dimerized
low-temperature structure, and one-particle spectral function of
the Hubbard model calculated by the DDMRG method ($U=3.3$\,eV,
$t=0.23$\,eV).}
\end{figure*}

The spectral changes in the $b$-axis ARPES data are somewhat more
complex (right panel of Fig.~\ref{Figure5}). They start out at
$\Gamma$ with a single peak at $\approx 1.5$\,eV below the
chemical potential. With increasing momentum the peak shifts
clearly towards $\mu_{exp}$ and reaches its smallest binding
energy about halfway between $\Gamma$ and the zone edge. For even
larger momentum the peak rapidly drops in intensity and seems to
move back to slightly higher binding energy. At the same time a
new feature appears at $\approx -2.5$\,eV until at the zone
boundary (Y) the spectral shape has evolved into a broad hump.
This behavior and the relatively large broadening of these
structures have been reproducibly observed on many different
samples and is hence to be taken as intrinsic.

Figure~\ref{Figure7} shows the experimental $b$-axis dispersion as
ARPES intensity plot in comparison with various theoretical
calculations. Starting with the theoretical LDA+U$^{\mathrm{FM}}$
dispersion calculated for the high-temperature (non-dimerized)
structure we find pronounced disagreement with the experiment,
particularly in the second half of the $\Gamma$Y direction where
the theoretical bands continue to disperse upwards, whereas the
experimental dispersion turns over and bends downwards until at
the Y-point the spectral weight distribution is strongly broadened
and seems to reach even higher binding energy than at $\Gamma$.
Using antiferromagnetic spin alignment in the LDA+U approach
doubles the unit cell and hence results in a dispersion symmetric
about $\frac{1}{2}\Gamma$Y (center panel of Fig.~\ref{Figure7}).
Although LDA+U$^{\mathrm{AFM}}$ thus reproduces the dispersion
maximum halfway along $\Gamma$Y, it does clearly not account for
the asymmetric experimental behavior and the strong broadening
towards the Y-point. Motivated by speculations that the unusual
high-temperature behavior of TiOCl could be caused by fluctuations
of the spin-Peierls order parameter, Fig.~\ref{Figure7} shows
LDA+U$^{\mathrm{AFM}}$ dispersions for the dimerized
low-temperature phase.\cite{Shaz05} We note that such fluctuation
effects have been observed in {\it charge} Peierls systems well
above the actual transition temperature.\cite{Schaefer01} However,
as is evident from the figure, the effect of (fluctuating)
dimerization results in band doubling but is otherwise rather
small and hence cannot explain the phenomenology of the ARPES
data.

The right panel of Fig.~\ref{Figure7} finally displays the
momentum-resolved spectral weight distribution of the 1D
single-band Hubbard model, calculated within DDMRG for the $U$ and
$t$ parameters of the LDA+U calculations. We note that in this
case the spectra are entirely of incoherent nature and correspond
to momentum-dependent continua, which are structured in intensity
due to the phase space available for decomposition of a real
(photo-)hole into separate collective spinon and holon
excitations. As a consequence a number of different dispersive
structures appear, whose detailed (non-quasiparticle) nature are
discussed in, {\it e.g.}, Refs.~\onlinecite{Voit95} and
\onlinecite{Benthien04}. Comparing the DDMRG result with our ARPES
data we find several corresponding features such as the initial
upward dispersion (due to spin and charge branches in the Hubbard
model \cite{Benthien04}), the dispersion maximum at
$\frac{1}{2}\Gamma$Y, and the asymmetric shift of weight to larger
binding energy towards the Y-point. Also the overall energy width
of the spectral weight distribution is greater than in the LDA+U
calculations. On the other hand, the experiment does not show the
pronounced spin-charge splitting between $\Gamma$ and
$\frac{1}{2}\Gamma$Y predicted by DDMRG, nor the holon ``shadow
band'' \cite{Benthien04} which in the Hubbard model spectrum
disperses downwards from $\Gamma$.

In addition, we calculated the spectral functions of two closely
related Hubbard-type models in the Mott-insulating phase: The
extended Hubbard model which includes a next-neighbor Coulomb
repulsion $V$ and the $t$-$t'$-$U$ model\cite{Daul00} that takes
into account a next-neighbor hopping $t'$. The resulting spectral
functions (not shown) for $V/t < 2$ and $t'/t<0.5$, respectively,
display no qualitative differences compared to the simple Hubbard
model spectral function. However, the spectral weight of the spin
branch in the $t$-$t'$-$U$ model is significantly reduced for
appropriate parameters ($U/t=15.7$,
$t'/t=0.14$)\cite{Saha-Dasgupta04}. This may indicate why the
spin-charge splitting is not resolved experimentally.

Overall, the agreement is much better for our Hubbard model
calculations than for the LDA+U band dispersions, suggesting that
the experimental spectra are indeed dominated by electronic
correlation effects. We note, however, that we cannot really
expect the 1D single-band Hubbard model to account fully for our
spectra, as it completely ignores the orbital degrees of freedom.
The crystal-field splitting between the $d_{xy}$ ground state and
the excited $d_{xz,yz}$ states is only a few
100\,meV\cite{Seidel03,Ruckamp05b} and hence virtual excitations
into these states will be even more important than the double
occupations on the same site (with energy scale $U=3.3$\,eV)
already contained in the single-band model. Therefore, it would be
highly desirable to compare our ARPES spectra to that of a
suitable multi-orbital 1D Hubbard model which however is still out
of reach for the DDMRG.

\subsection{Polarization-dependent photoemission}

As already discussed in the introduction, the high-temperature
phase of TiOCl is characterized by anomalous broadening of the
phonon lines in Raman and infrared
spectroscopy.\cite{Lemmens04,Caimi04} This indicates strong
coupling of the lattice to the electronic (spin or orbital)
degrees of freedom. Indeed, LDA+U calculations for frozen lattice
distortions corresponding to the relevant phonon modes have shown
that the orbital ground state may switch from $3d_{xy}$ to
$3d_{xz,yz}$ for sufficiently large
amplitude.\cite{Saha-Dasgupta04} Such a dynamical Jahn-Teller
effect would lead to an orbitally mixed character of the
time-averaged ground state.

\begin{figure}
\includegraphics[width=8cm]{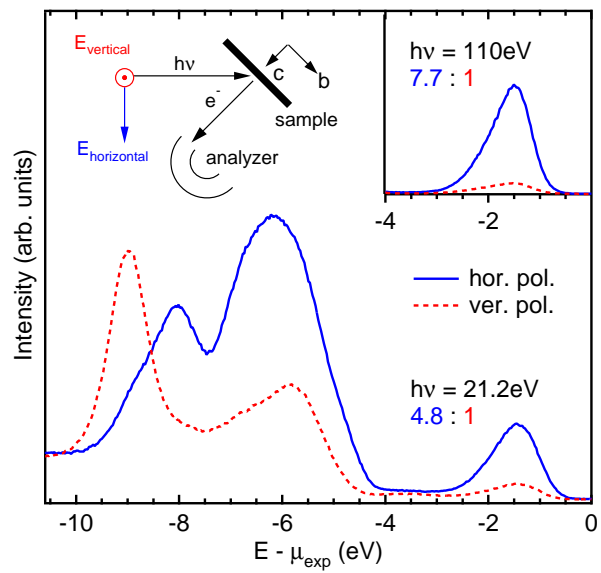}
\caption{\label{Figure8}(Color online) Photoemission spectra
measured at the $\Gamma$-point with horizontal and vertical light
polarization, respectively (T\,=\,300\,K). The inset shows the
corresponding photoemission spectra measured at another photon
energy (T\,=\,365\,K). The experimental setup is sketched in the
upper left corner.}
\end{figure}

Experimental information on orbital symmetry can be obtained from
polarization-dependent ARPES, making use of selection rules
realized for special experimental geometries. Assume the direction
of the incident linearly polarized light and the emission
direction of the photoelectrons lie within the same crystal mirror
plane. Then with the polarization vector within and perpendicular
to the mirror plane the ejected electrons can stem only from
states with well-defined even or odd parity with respect to this
plane, respectively.\cite{Damascelli03} If we choose the
experimental setup as sketched in Fig.~\ref{Figure8} the
$d_{xy}$-derived band states of TiOCl are even with respect to the
($b$,$c$) mirror plane while the $d_{xz,yz}$ states are odd. With
the ($b$,$c$)-plane lying horizontally, photoemission from the
$d_{xy}$ states is hence dipole-allowed only for horizontal light
polarization, whereas $d_{xz,yz}$-emission can be observed for
vertical polarization only.

In the main panel of Fig.~\ref{Figure8} we display the results of
such an ARPES experiment obtained with 21.2\,eV photons at room
temperature. The spectra were recorded at normal emission ({\it
i.e.}, at the $\Gamma$ point) with horizontal (blue) and vertical
(red) polarization, respectively. Since a laboratory He lamp with
a rotatable polarizer was used, the photon flux is not changed by
switching the polarization and the spectra can be normalized to
equal integration times per channel. At first glance, one can see
that the intensity distribution over the whole valence band indeed
is strongly affected by polarization effects. Focussing on the
Ti~3$d$ states reaching down to about 3.5\,eV below ${\mu}_{exp}$
their spectral weight is significantly reduced but not completely
suppressed upon switching the polarization from horizontal to
vertical, indicating the dominance of $d_{xy}$ emission. The
residual $1:4.8$ weight for vertical polarization can be
quantitatively accounted for by the finite degree of light
polarization ($\approx 85$\,\%), possible small sample
misalignment and the effect of thermally activated
symmetry-breaking phonons.\cite{footnote-phonons} Similar data for
the Ti~3$d$ derived part of the electronic structure using
polarized synchrotron radiation is depicted in the inset of
Fig.~\ref{Figure8}. These spectra were recorded at a photon energy
of 110\,eV and a temperature $T=365$\,K and normalized to equal
photon flux. Due to the higher polarization degree of the
synchrotron radiation the residual relative weight for vertical
polarization is even further reduced. We note that this
contribution does not vary down to room temperature within
experimental accuracy again indicating that phonon-induced orbital
fluctuations are not important in this temperature range.

Hence we have direct experimental evidence that there is no
sizable dynamical Jahn-Teller admixture of $d_{xz}$ and $d_{yz}$
states to the $d_{xy}$ ground state at room temperature. This is
in line with the reasoning based on recent cluster calculations
combined with polarization-dependent optical data\cite{Ruckamp05},
although such an effect cannot be excluded for far higher
temperatures.

\section{Conclusions}
%
%

In summary, we have investigated the electronic structure of TiOCl
by means of angle-resolved photoelectron spectroscopy at and
slightly above room temperature. From those measurements we
confirm the predicted quasi-one-dimensional nature of the
low-lying energy excitations. However, the electron dispersions
along the one-dimensional axis cannot be explained by LDA+U or
DDMRG calculations of the one-dimensional single-band Hubbard
model. From the specific polarization dependence of the Ti~3$d$
spectral weight close to the chemical potential sizeable
phonon-induced admixtures of $d_{xz}$ and $d_{yz}$ derived states
to the $d_{xy}$ ground state can be ruled out. Thus a scenario
where the non-canonical behavior of the spin-Peierls transition in
this compound is traced back to orbital fluctuations within a
dynamical Jahn-Teller scenario does not seem plausible.

\begin{acknowledgments}
We are indebted to V.~N.~Strocov and L.~Patthey for their support
at the Swiss Light Source and to M.~Knupfer and S.~Borisenko for
the loan of their experimental end station. We would also like to
thank H.~Rosner for useful discussions and A.~Hartmann for
technical assistance. This work was supported by the Deutsche
Forschungsgemeinschaft through SFB 484 and grant CL 124/3-3.
\end{acknowledgments}


\begin{thebibliography}{31}
\expandafter\ifx\csname
natexlab\endcsname\relax\def\natexlab#1{#1}\fi
\expandafter\ifx\csname bibnamefont\endcsname\relax
  \def\bibnamefont#1{#1}\fi
\expandafter\ifx\csname bibfnamefont\endcsname\relax
  \def\bibfnamefont#1{#1}\fi
\expandafter\ifx\csname citenamefont\endcsname\relax
  \def\citenamefont#1{#1}\fi
\expandafter\ifx\csname url\endcsname\relax
  \def\url#1{\texttt{#1}}\fi
\expandafter\ifx\csname
urlprefix\endcsname\relax\def\urlprefix{URL }\fi
\providecommand{\bibinfo}[2]{#2}
\providecommand{\eprint}[2][]{\url{#2}}

\bibitem[{\citenamefont{Imada et~al.}(1999)\citenamefont{Imada, Fujimori, and
  Tokura}}]{Imada99}
\bibinfo{author}{\bibfnamefont{M.}~\bibnamefont{Imada}},
  \bibinfo{author}{\bibfnamefont{A.}~\bibnamefont{Fujimori}}, \bibnamefont{and}
  \bibinfo{author}{\bibfnamefont{Y.}~\bibnamefont{Tokura}},
  \bibinfo{journal}{Rev. Mod. Phys.} \textbf{\bibinfo{volume}{70}},
  \bibinfo{pages}{1039} (\bibinfo{year}{1999}).

\bibitem[{\citenamefont{Dagotto}(1999)}]{Dagotto99}
\bibinfo{author}{\bibfnamefont{E.}~\bibnamefont{Dagotto}},
  \bibinfo{journal}{Rep. Prog. Phys.} \textbf{\bibinfo{volume}{62}},
  \bibinfo{pages}{1525} (\bibinfo{year}{1999}).

\bibitem[{\citenamefont{Voit}(1995)}]{Voit95}
\bibinfo{author}{\bibfnamefont{J.}~\bibnamefont{Voit}}, \bibinfo{journal}{Rep.
  Prog. Phys.} \textbf{\bibinfo{volume}{58}}, \bibinfo{pages}{977}
  (\bibinfo{year}{1995}).

\bibitem[{\citenamefont{Sch{\"a}fer et~al.}(1958)\citenamefont{Sch{\"a}fer,
  Wartenpfuhl, and Weise}}]{Schaefer58}
\bibinfo{author}{\bibfnamefont{H.}~\bibnamefont{Sch{\"a}fer}},
  \bibinfo{author}{\bibfnamefont{F.}~\bibnamefont{Wartenpfuhl}},
  \bibnamefont{and} \bibinfo{author}{\bibfnamefont{E.}~\bibnamefont{Weise}},
  \bibinfo{journal}{Z. Anorg. Allg. Chemie} \textbf{\bibinfo{volume}{295}},
  \bibinfo{pages}{268} (\bibinfo{year}{1958}).

\bibitem[{\citenamefont{Seidel et~al.}(2003)\citenamefont{Seidel, Marianetti,
  Chou, Ceder, and Lee}}]{Seidel03}
\bibinfo{author}{\bibfnamefont{A.}~\bibnamefont{Seidel}},
  \bibinfo{author}{\bibfnamefont{C.~A.} \bibnamefont{Marianetti}},
  \bibinfo{author}{\bibfnamefont{F.~C.} \bibnamefont{Chou}},
  \bibinfo{author}{\bibfnamefont{G.}~\bibnamefont{Ceder}}, \bibnamefont{and}
  \bibinfo{author}{\bibfnamefont{P.~A.} \bibnamefont{Lee}},
  \bibinfo{journal}{Phys. Rev. B} \textbf{\bibinfo{volume}{67}},
  \bibinfo{pages}{020405(R)} (\bibinfo{year}{2003}).

\bibitem[{\citenamefont{Saha-Dasgupta et~al.}(2004)\citenamefont{Saha-Dasgupta,
  Valent\'i, Rosner, and Gros}}]{Saha-Dasgupta04}
\bibinfo{author}{\bibfnamefont{T.}~\bibnamefont{Saha-Dasgupta}},
  \bibinfo{author}{\bibfnamefont{R.}~\bibnamefont{Valent\'i}},
  \bibinfo{author}{\bibfnamefont{H.}~\bibnamefont{Rosner}}, \bibnamefont{and}
  \bibinfo{author}{\bibfnamefont{C.}~\bibnamefont{Gros}},
  \bibinfo{journal}{Europhys. Lett.} \textbf{\bibinfo{volume}{67}},
  \bibinfo{pages}{63} (\bibinfo{year}{2004}).

\bibitem[{foo({\natexlab{a}})}]{footnote-xyz}
\bibinfo{note}{For the orientation of the orbitals we use the notation of
  Ref.~\onlinecite{Saha-Dasgupta04}.}

\bibitem[{\citenamefont{Shaz et~al.}(2005)\citenamefont{Shaz, van Smaalen,
  Palatinus, Hoinkis, Klemm, Horn, and Claessen}}]{Shaz05}
\bibinfo{author}{\bibfnamefont{M.}~\bibnamefont{Shaz}},
  \bibinfo{author}{\bibfnamefont{S.}~\bibnamefont{van Smaalen}},
  \bibinfo{author}{\bibfnamefont{L.}~\bibnamefont{Palatinus}},
  \bibinfo{author}{\bibfnamefont{M.}~\bibnamefont{Hoinkis}},
  \bibinfo{author}{\bibfnamefont{M.}~\bibnamefont{Klemm}},
  \bibinfo{author}{\bibfnamefont{S.}~\bibnamefont{Horn}}, \bibnamefont{and}
  \bibinfo{author}{\bibfnamefont{R.}~\bibnamefont{Claessen}},
  \bibinfo{journal}{Phys. Rev. B} \textbf{\bibinfo{volume}{71}},
  \bibinfo{pages}{100405(R)} (\bibinfo{year}{2005}).

\bibitem[{\citenamefont{Imai and Chou}()}]{Imai03}
\bibinfo{author}{\bibfnamefont{T.}~\bibnamefont{Imai}} \bibnamefont{and}
  \bibinfo{author}{\bibfnamefont{F.~C.} \bibnamefont{Chou}},
  \eprint{cond-mat/0301425}.

\bibitem[{\citenamefont{Kataev et~al.}(2003)\citenamefont{Kataev, Baier,
  M{\"o}ller, Jongen, Meyer, and Freimuth}}]{Kataev03}
\bibinfo{author}{\bibfnamefont{V.}~\bibnamefont{Kataev}},
  \bibinfo{author}{\bibfnamefont{J.}~\bibnamefont{Baier}},
  \bibinfo{author}{\bibfnamefont{A.}~\bibnamefont{M{\"o}ller}},
  \bibinfo{author}{\bibfnamefont{L.}~\bibnamefont{Jongen}},
  \bibinfo{author}{\bibfnamefont{G.}~\bibnamefont{Meyer}}, \bibnamefont{and}
  \bibinfo{author}{\bibfnamefont{A.}~\bibnamefont{Freimuth}},
  \bibinfo{journal}{Phys. Rev. B} \textbf{\bibinfo{volume}{68}},
  \bibinfo{pages}{140405(R)} (\bibinfo{year}{2003}).

\bibitem[{\citenamefont{Lemmens et~al.}(2004)\citenamefont{Lemmens, Choi,
  Caimi, Degiorgi, Kovaleva, Seidel, and Chou}}]{Lemmens04}
\bibinfo{author}{\bibfnamefont{P.}~\bibnamefont{Lemmens}},
  \bibinfo{author}{\bibfnamefont{K.~Y.} \bibnamefont{Choi}},
  \bibinfo{author}{\bibfnamefont{G.}~\bibnamefont{Caimi}},
  \bibinfo{author}{\bibfnamefont{L.}~\bibnamefont{Degiorgi}},
  \bibinfo{author}{\bibfnamefont{N.~N.} \bibnamefont{Kovaleva}},
  \bibinfo{author}{\bibfnamefont{A.}~\bibnamefont{Seidel}}, \bibnamefont{and}
  \bibinfo{author}{\bibfnamefont{F.~C.} \bibnamefont{Chou}},
  \bibinfo{journal}{Phys. Rev. B} \textbf{\bibinfo{volume}{70}},
  \bibinfo{pages}{134429} (\bibinfo{year}{2004}).

\bibitem[{\citenamefont{Caimi et~al.}(2004)\citenamefont{Caimi, Degiorgi,
  Kovaleva, Lemmens, and Chou}}]{Caimi04}
\bibinfo{author}{\bibfnamefont{G.}~\bibnamefont{Caimi}},
  \bibinfo{author}{\bibfnamefont{L.}~\bibnamefont{Degiorgi}},
  \bibinfo{author}{\bibfnamefont{N.~N.} \bibnamefont{Kovaleva}},
  \bibinfo{author}{\bibfnamefont{P.}~\bibnamefont{Lemmens}}, \bibnamefont{and}
  \bibinfo{author}{\bibfnamefont{F.~C.} \bibnamefont{Chou}},
  \bibinfo{journal}{Phys. Rev. B} \textbf{\bibinfo{volume}{69}},
  \bibinfo{pages}{125108} (\bibinfo{year}{2004}).

\bibitem[{\citenamefont{Hemberger et~al.}()\citenamefont{Hemberger, Hoinkis,
  Klemm, Sing, Claessen, Horn, and Loidl}}]{Hemberger05}
\bibinfo{author}{\bibfnamefont{J.}~\bibnamefont{Hemberger}},
  \bibinfo{author}{\bibfnamefont{M.}~\bibnamefont{Hoinkis}},
  \bibinfo{author}{\bibfnamefont{M.}~\bibnamefont{Klemm}},
  \bibinfo{author}{\bibfnamefont{M.}~\bibnamefont{Sing}},
  \bibinfo{author}{\bibfnamefont{R.}~\bibnamefont{Claessen}},
  \bibinfo{author}{\bibfnamefont{S.}~\bibnamefont{Horn}}, \bibnamefont{and}
  \bibinfo{author}{\bibfnamefont{A.}~\bibnamefont{Loidl}},
  \eprint{cond-mat/0501517}.

\bibitem[{\citenamefont{R{\"u}ckamp
  et~al.}({\natexlab{a}})\citenamefont{R{\"u}ckamp, Baier, Kriener, Haverkort,
  Lorenz, Uhrig, Jongen, M{\"o}ller, Meyer, and Gr{\"u}ninger}}]{Ruckamp05}
\bibinfo{author}{\bibfnamefont{R.}~\bibnamefont{R{\"u}ckamp}},
  \bibinfo{author}{\bibfnamefont{J.}~\bibnamefont{Baier}},
  \bibinfo{author}{\bibfnamefont{M.}~\bibnamefont{Kriener}},
  \bibinfo{author}{\bibfnamefont{M.}~\bibnamefont{Haverkort}},
  \bibinfo{author}{\bibfnamefont{T.}~\bibnamefont{Lorenz}},
  \bibinfo{author}{\bibfnamefont{G.}~\bibnamefont{Uhrig}},
  \bibinfo{author}{\bibfnamefont{L.}~\bibnamefont{Jongen}},
  \bibinfo{author}{\bibfnamefont{A.}~\bibnamefont{M{\"o}ller}},
  \bibinfo{author}{\bibfnamefont{G.}~\bibnamefont{Meyer}}, \bibnamefont{and}
  \bibinfo{author}{\bibfnamefont{M.}~\bibnamefont{Gr{\"u}ninger}},
  \eprint{cond-mat/0503409}.

\bibitem[{\citenamefont{Palatinus et~al.}(2005)\citenamefont{Palatinus,
  Sch{\"o}nleber, and van Smaalen}}]{vanSmaalen05}
\bibinfo{author}{\bibfnamefont{L.}~\bibnamefont{Palatinus}},
  \bibinfo{author}{\bibfnamefont{A.}~\bibnamefont{Sch{\"o}nleber}},
  \bibnamefont{and} \bibinfo{author}{\bibfnamefont{S.}~\bibnamefont{van
  Smaalen}}, \bibinfo{journal}{Acta Crystallographica Section C}
  \textbf{\bibinfo{volume}{61}}, \bibinfo{pages}{i47} (\bibinfo{year}{2005}).

\bibitem[{\citenamefont{Beynon and Wilson}(1993)}]{Beynon93}
\bibinfo{author}{\bibfnamefont{R.~J.} \bibnamefont{Beynon}} \bibnamefont{and}
  \bibinfo{author}{\bibfnamefont{J.~A.} \bibnamefont{Wilson}},
  \bibinfo{journal}{J. Phys.: Condens. Matter} \textbf{\bibinfo{volume}{5}},
  \bibinfo{pages}{1983} (\bibinfo{year}{1993}).

\bibitem[{\citenamefont{Craco et~al.}()\citenamefont{Craco, Laad, and
  M{\"u}ller-Hartmann}}]{Craco04}
\bibinfo{author}{\bibfnamefont{L.}~\bibnamefont{Craco}},
  \bibinfo{author}{\bibfnamefont{M.~S.} \bibnamefont{Laad}}, \bibnamefont{and}
  \bibinfo{author}{\bibfnamefont{E.}~\bibnamefont{M{\"u}ller-Hartmann}},
  \eprint{cond-mat/0410472}.

\bibitem[{\citenamefont{Perdew et~al.}(1996)\citenamefont{Perdew, Burke, and
  Ernzerhof}}]{Perdew96}
\bibinfo{author}{\bibfnamefont{J.}~\bibnamefont{Perdew}},
  \bibinfo{author}{\bibfnamefont{S.}~\bibnamefont{Burke}}, \bibnamefont{and}
  \bibinfo{author}{\bibfnamefont{M.}~\bibnamefont{Ernzerhof}},
  \bibinfo{journal}{Phys. Rev. Lett.} \textbf{\bibinfo{volume}{77}},
  \bibinfo{pages}{3865} (\bibinfo{year}{1996}).

\bibitem[{\citenamefont{Anisimov et~al.}(1993)\citenamefont{Anisimov, Solovyev,
  Korotin, Czyzyk, and Sawatzky}}]{Anisimov93}
\bibinfo{author}{\bibfnamefont{V.~I.} \bibnamefont{Anisimov}},
  \bibinfo{author}{\bibfnamefont{I.~V.} \bibnamefont{Solovyev}},
  \bibinfo{author}{\bibfnamefont{M.~A.} \bibnamefont{Korotin}},
  \bibinfo{author}{\bibfnamefont{M.~T.} \bibnamefont{Czyzyk}},
  \bibnamefont{and} \bibinfo{author}{\bibfnamefont{G.~A.}
  \bibnamefont{Sawatzky}}, \bibinfo{journal}{Phys. Rev. B}
  \textbf{\bibinfo{volume}{48}}, \bibinfo{pages}{16929} (\bibinfo{year}{1993}).

\bibitem[{\citenamefont{Blaha et~al.}(2001)\citenamefont{Blaha, Schwarz, and
  Luitz}}]{wien2k}
\bibinfo{author}{\bibfnamefont{P.}~\bibnamefont{Blaha}},
  \bibinfo{author}{\bibfnamefont{K.}~\bibnamefont{Schwarz}}, \bibnamefont{and}
  \bibinfo{author}{\bibfnamefont{J.}~\bibnamefont{Luitz}}
  (\bibinfo{year}{2001}), \bibinfo{note}{computer code WIEN2k, Techn. Univ.
  Wien, Vienna}, \urlprefix\url{http://www.wien2k.at}.

\bibitem[{\citenamefont{Jeckelmann}(2002)}]{Jeckelmann02}
\bibinfo{author}{\bibfnamefont{E.}~\bibnamefont{Jeckelmann}},
  \bibinfo{journal}{Rev. Mod. Phys.} \textbf{\bibinfo{volume}{66}},
  \bibinfo{pages}{045114} (\bibinfo{year}{2002}).

\bibitem[{\citenamefont{Benthien et~al.}(2004)\citenamefont{Benthien, Gebhard,
  and Jeckelmann}}]{Benthien04}
\bibinfo{author}{\bibfnamefont{H.}~\bibnamefont{Benthien}},
  \bibinfo{author}{\bibfnamefont{F.}~\bibnamefont{Gebhard}}, \bibnamefont{and}
  \bibinfo{author}{\bibfnamefont{E.}~\bibnamefont{Jeckelmann}},
  \bibinfo{journal}{Phys. Rev. Lett.} \textbf{\bibinfo{volume}{92}},
  \bibinfo{pages}{256401} (\bibinfo{year}{2004}).

\bibitem[{\citenamefont{Maule et~al.}(1988)\citenamefont{Maule, Tothill,
  Strange, and Wilson}}]{Maule88}
\bibinfo{author}{\bibfnamefont{C.~H.} \bibnamefont{Maule}},
  \bibinfo{author}{\bibfnamefont{J.~N.} \bibnamefont{Tothill}},
  \bibinfo{author}{\bibfnamefont{P.}~\bibnamefont{Strange}}, \bibnamefont{and}
  \bibinfo{author}{\bibfnamefont{J.~A.} \bibnamefont{Wilson}},
  \bibinfo{journal}{J. Phys. C: Solid State Phys.}
  \textbf{\bibinfo{volume}{21}}, \bibinfo{pages}{2153} (\bibinfo{year}{1988}).

\bibitem[{\citenamefont{Gr{\"u}ninger}()}]{Grueninger04}
\bibinfo{author}{\bibfnamefont{M.}~\bibnamefont{Gr{\"u}ninger}},
  \bibinfo{howpublished}{private communication}.

\bibitem[{\citenamefont{Saha-Dasgupta et~al.}(2005)\citenamefont{Saha-Dasgupta,
  Lichtenstein, and Valent\'i}}]{Saha-Dasgupta05}
\bibinfo{author}{\bibfnamefont{T.}~\bibnamefont{Saha-Dasgupta}},
  \bibinfo{author}{\bibfnamefont{A.}~\bibnamefont{Lichtenstein}},
  \bibnamefont{and}
  \bibinfo{author}{\bibfnamefont{R.}~\bibnamefont{Valent\'i}},
  \bibinfo{journal}{Phys. Rev. B} \textbf{\bibinfo{volume}{71}},
  \bibinfo{pages}{153108} (\bibinfo{year}{2005}).

\bibitem[{\citenamefont{Lichtenstein and Katsnelson}(1998)}]{Lichtenstein98}
\bibinfo{author}{\bibfnamefont{A.~I.} \bibnamefont{Lichtenstein}}
  \bibnamefont{and} \bibinfo{author}{\bibfnamefont{M.~I.}
  \bibnamefont{Katsnelson}}, \bibinfo{journal}{Phys. Rev. B}
  \textbf{\bibinfo{volume}{57}}, \bibinfo{pages}{6884} (\bibinfo{year}{1998}).

\bibitem[{\citenamefont{Sch{\"a}fer et~al.}(2001)\citenamefont{Sch{\"a}fer,
  Rotenberg, Kevan, Blaha, Claessen, and Thorne}}]{Schaefer01}
\bibinfo{author}{\bibfnamefont{J.}~\bibnamefont{Sch{\"a}fer}},
  \bibinfo{author}{\bibfnamefont{E.}~\bibnamefont{Rotenberg}},
  \bibinfo{author}{\bibfnamefont{S.~D.} \bibnamefont{Kevan}},
  \bibinfo{author}{\bibfnamefont{P.}~\bibnamefont{Blaha}},
  \bibinfo{author}{\bibfnamefont{R.}~\bibnamefont{Claessen}}, \bibnamefont{and}
  \bibinfo{author}{\bibfnamefont{R.~E.} \bibnamefont{Thorne}},
  \bibinfo{journal}{Phys. Rev. Lett.} \textbf{\bibinfo{volume}{87}},
  \bibinfo{pages}{196403} (\bibinfo{year}{2001}).

\bibitem[{\citenamefont{Daul and Noack}(2000)}]{Daul00}
\bibinfo{author}{\bibfnamefont{S.}~\bibnamefont{Daul}} \bibnamefont{and}
  \bibinfo{author}{\bibfnamefont{R.}~\bibnamefont{Noack}},
  \bibinfo{journal}{Phys. Rev. B} \textbf{\bibinfo{volume}{61}},
  \bibinfo{pages}{1646} (\bibinfo{year}{2000}).

\bibitem[{\citenamefont{R{\"u}ckamp
  et~al.}({\natexlab{b}})\citenamefont{R{\"u}ckamp, Benckiser, Haverkort, Roth,
  Lorenz, Freimuth, Jongen, M{\"o}ller, Meyer, Reutler et~al.}}]{Ruckamp05b}
\bibinfo{author}{\bibfnamefont{R.}~\bibnamefont{R{\"u}ckamp}},
  \bibinfo{author}{\bibfnamefont{E.}~\bibnamefont{Benckiser}},
  \bibinfo{author}{\bibfnamefont{M.}~\bibnamefont{Haverkort}},
  \bibinfo{author}{\bibfnamefont{H.}~\bibnamefont{Roth}},
  \bibinfo{author}{\bibfnamefont{T.}~\bibnamefont{Lorenz}},
  \bibinfo{author}{\bibfnamefont{A.}~\bibnamefont{Freimuth}},
  \bibinfo{author}{\bibfnamefont{L.}~\bibnamefont{Jongen}},
  \bibinfo{author}{\bibfnamefont{A.}~\bibnamefont{M{\"o}ller}},
  \bibinfo{author}{\bibfnamefont{G.}~\bibnamefont{Meyer}},
  \bibinfo{author}{\bibfnamefont{P.}~\bibnamefont{Reutler}},
  \bibnamefont{et~al.}, \eprint{cond-mat/0503405}.

\bibitem[{\citenamefont{Damascelli et~al.}(2003)\citenamefont{Damascelli,
  Hussain, and Shen}}]{Damascelli03}
\bibinfo{author}{\bibfnamefont{A.}~\bibnamefont{Damascelli}},
  \bibinfo{author}{\bibfnamefont{Z.}~\bibnamefont{Hussain}}, \bibnamefont{and}
  \bibinfo{author}{\bibfnamefont{Z.-X.} \bibnamefont{Shen}},
  \bibinfo{journal}{Rev. Mod. Phys.} \textbf{\bibinfo{volume}{75}},
  \bibinfo{pages}{473} (\bibinfo{year}{2003}).

\bibitem[{foo({\natexlab{b}})}]{footnote-phonons}
\bibinfo{note}{Note that the phonons involved in the dynamical Jahn-Teller
  effect discussed in Ref.~\onlinecite{Saha-Dasgupta04} {\it do not} break the
  mirror symmetry of the ($b$,$c$)-plane.}

\end{thebibliography}
\end{document}